\documentclass[twocolumn,aps,pra,showpacs,preprintnumbers,floatfix]{revtex4}
\usepackage[latin9]{inputenc}
\usepackage[active]{srcltx}
\usepackage{amsmath}
\usepackage{amssymb}
\usepackage{graphicx}
\usepackage{esint}

\makeatletter
\@ifundefined{textcolor}{}
{
 \definecolor{BLACK}{gray}{0}
 \definecolor{WHITE}{gray}{1}
 \definecolor{RED}{rgb}{1,0,0}
 \definecolor{GREEN}{rgb}{0,1,0}
 \definecolor{BLUE}{rgb}{0,0,1}
 \definecolor{CYAN}{cmyk}{1,0,0,0}
 \definecolor{MAGENTA}{cmyk}{0,1,0,0}
 \definecolor{YELLOW}{cmyk}{0,0,1,0}
}
\makeatother

\begin{document}

\title{Quantum phenomena in a chirped parametric anharmonic oscillator}
\author{I. Barth and L. Friedland}
\affiliation{Racah Institute of Physics, Hebrew University of Jerusalem, Jerusalem 91904,
Israel}

\begin{abstract}
The parametric ladder climbing (successive Landau-Zener-type transitions)
and the quantum saturation of the threshold for the classical parametric
autoresonance due to the zero point fluctuations at low temperatures are
discussed. The probability for capture into the chirped parametric resonance
is found by solving the Schrodinger equation in the energy basis and the
associated resonant phase space dynamics is illustrated via the Wigner
distribution. The numerical threshold for the efficient capture into the
resonance is compared with the classical and quantum theories in different
parameter regimes.
\end{abstract}

\maketitle

\emph{Introduction.-} The parametric resonance is one of the most
interesting and frequently used phenomena in classical and quantum
dynamics. It occurs when the natural frequency of a system depends on a
parameter oscillating (modulated) at twice the natural system's frequency
\cite{Landau_Mechanics,Leonardo,Braun,Dykman2011,Gevorgyan}. In the well
studied stationary case, the modulation frequency is constant. However, in
nonlinear systems the stationary parametric amplification is restricted to
small amplitudes, since at larger amplitudes the resonance (phase-locking)
is destroyed due to the nonlinear frequency shift \cite{Landau_Mechanics}. A
robust method to overcome this limitation is to slowly vary the modulation
frequency so that the resonance condition is preserved despite the increase
of the amplitude of oscillations. This nonstationary phenomenon is called
\emph{parametric autoresonance }(PAR). The PAR was studied in such classical
oscillatory systems as the anharmonic oscillator \cite{Khain,Kiselev},
Faraday waves \cite{Assaf,Ben-David}, and plasmas \cite{lazar90}. A related,
but different control method is the direct autoresonance (AR), where instead
of parametric modulations, a chirped \emph{external} driving force is
applied. The direct AR was extensively studied and implemented in various
classical and quantum physical systems \cite{ALPHA
PRL,Barak,lazar121,lazar90,Manfredi 2007,Danielson}.

When studying the classical to quantum transitions in the direct
chirped-driven oscillator, one identifies two limits, where quantum
dynamical effects are significant. The first is the saturation of
temperature-dependent classical observables at small temperatures due to the
zero point quantum fluctuations \cite{ido4,ido5}. In the second limit, at
sufficiently large anharmonicity, the smooth classical AR dynamics of many
simultaneously coupled energy levels transforms into a quantum ladder
climbing involving successive two-level Landau-Zener (LZ) transitions \cite%
{LZ,Marcus_theory,ido5,ido6}. These two quantum limits were also studied
recently in the case of the direct chirped subharmonic (two-photon)
resonance \cite{ido7}. This letter, for the \ first time, discusses
the analogous quantum phenomena in application to the PAR.

\emph{The model.-} The simplest system exhibiting nonlinear parametric
resonance is the anharmonic oscillator governed by Hamiltonian
\begin{equation}
H=\frac{p^{2}}{2}+\left( 1+\varepsilon \cos \varphi \right) \frac{x^{2}}{2}%
+\beta \frac{x^{4}}{4}  \label{eq:Hamiltonian}
\end{equation}%
(here and below all variables and parameters are dimensionless). The
frequency of the modulation is chirped, $\omega \equiv d\varphi /dt=2+\alpha
t$, passing the linear resonance at $t=0.$ We expand the wave function of
the oscillator, $|\psi \rangle =\sum_{n}c_{n}|\psi _{n}\rangle $, in the
energy basis $|\psi _{n}\rangle $ of the unmodulated Hamiltonian i.e., $%
H(\varepsilon =0)|\psi _{n}\rangle =E_{n}|\psi _{n}\rangle \ $and $\langle
\psi _{k}|\psi _{n}\rangle =\delta _{k,n}$. In this basis, the dimensionless
($\hbar =1$) Schrodinger equation is
\begin{equation}
i\frac{dc_{n}}{dt}=E_{n}c_{n}+\frac{\varepsilon }{2}\sum_{k}c_{k}\langle
\psi _{k}|\hat{x}^{2}|\psi _{n}\rangle \cos \varphi ,  \label{eq:schrodinger}
\end{equation}%
where for small $\beta $, the energy levels can be approximated as \cite%
{Landau QM}
\begin{equation}
E_{n}\approx n+\frac{1}{2}+\gamma (n^{2}+n)+\frac{3}{16}\beta ,  \label{En}
\end{equation}%
$n=0,1,2,...$, and $\gamma =\frac{3}{8}\beta $. We also assume weak
coupling, $\varepsilon \ll 1$ , and, consequently, neglect the nonlinear
corrections of order $\beta $ in the coupling term, $\langle \psi _{k}|\hat{x%
}^{2}|\psi _{n}\rangle \approx \frac{1}{2}[\sqrt{Q_{n}}\delta
_{k,n-2}+\left( 2n+1\right) \delta _{k,n}+\sqrt{Q_{n+1}}\delta _{k,n+2}]$,
where $Q_{n}=n(n+1)$. The resulting equation for $c_{n}$ is
\begin{eqnarray}
i\frac{dc_{n}}{dt} &=&E_{n}c_{n}+\frac{\varepsilon }{4}(\sqrt{Q_{n-1}}c_{n-2}+(2n+1)c_{n}  \notag \\
&+&\sqrt{Q_{n+1}}c_{n+2})\cos \varphi.
\label{eq:c_n}
\end{eqnarray}%
\begin{figure}[tb]
\includegraphics[width=9cm]{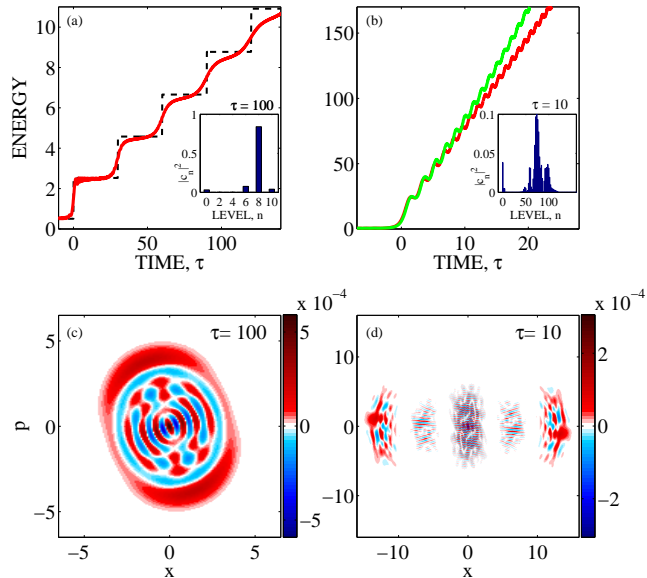}
\caption{(color online) The energy of the chirped parametric
oscillator in the quantum PLC (a) and the classical PAR (b) regimes versus
the slow time $\protect\tau$. In (b), the Schrodinger simulation (red line) is
compared to the classical solution (green
line). The inserts are snapshots of the quantum state in the energy basis at given times. The corresponding
snapshots of the Wigner distributions are shown in (c) and (d).}
\label{Flo:fig1}
\end{figure}

\emph{Numerical examples}.- At this stage, we illustrate the PLC and the PAR
in simulations. We have solved Eq. (\ref{eq:c_n}) numerically, subject to
ground state initial conditions, $c_{n}\left( t_{0}=-10/\sqrt{\alpha }%
\right) =\delta _{n,0}$, for two sets of parameters representing the quantum
PLC (see Fig. \ref{Flo:fig1}(a,c)) and the classical PAR (Fig. \ref{Flo:fig1}%
(b,d)). In the first example, $\{\alpha ,\beta ,\varepsilon
\}=\{10^{-6},0.01,0.04\}$ and $40$ energy levels are included in
simulations. The energy of the system versus the slow time $\tau =\sqrt{%
\alpha }t$ is shown in Fig. \ref{Flo:fig1}(a). One can see that the response
of the quantum anharmonic oscillator to the chirped parametric modulation
involves successive transitions between neighboring \emph{even} energy
levels, i.e., the PLC. We define the anharmonicity parameter $P_{2}=2\gamma /%
\sqrt{\alpha }$ \cite{Marcus_theory,ido5} ($P_{2}=10$ in this example) and
observe that the successive $n\rightarrow n+2$ transitions occur at times $%
\tau _{n}=4nP_{2}$, in agreement with the theory below (dashed black line).
To our knowledge, such ladder climbing dynamics in the parametrically modulated
anharmonic oscillator is observed for the first time.

The second example
(shown in Figs. \ref{Flo:fig1}(b,d)) uses the same initial conditions, but
$\{\alpha ,\beta ,\varepsilon \}=\{10^{-4},10^{-3},0.04\}$ and $%
250$ levels. Here, $P_{2}=0.075$ corresponding to the classical limit \cite%
{Marcus_theory,ido5}, where the dynamics involves many energy levels and the
energy grows as expected in the classical PAR \cite{Khain,lazar90}. We
compare this example with the classical simulations for the same
Hamiltonian (\ref{eq:Hamiltonian}), i.e. solve
\begin{equation}
d^{2}x/dt^{2}+(1+\varepsilon \cos \varphi )x+\beta x^{3}=0
\label{classical}
\end{equation}%
with the same parameters. The unique characteristic of the classical parametric
resonance is the existence of an unstable fixed point at zero energy \cite%
{Landau_Mechanics}. Therefore, the chirped excitation in the classical case
must involve nonzero initial conditions, e.g., a finite energy with random
phases, as considered in Ref. \cite{lazar90}, or a thermal distribution of
initial conditions, i.e., $f(x_{0},u_{0})=(2\pi T)^{-1}\exp \left[
-(x_{0}^{2}+u_{0}^{2})/(2T)\right] $, where $u=dx/dt$. The latter choice is
more suitable for studying the classical-quantum correspondence. We use $%
T=0.5$ associated with the energy of the quantum mechanical ground state of
the unmodulated linearized system in these simulations. The classical
averaged energy over $1000$ realizations is plotted in green in Fig. \ref{Flo:fig1}(b) showing a good
agreement with the quantum simulations. The deviation at large times is due
to higher order corrections of the energy levels not included in Eqs. (\ref%
{En}) and (\ref{eq:c_n}). The fact that the classical results can be
reconstructed by solving the quantum equations implies that the
correspondence principle is satisfied in the limit of small anharmonicity $%
(\beta \ll 1)$, where many energy levels are coupled simultaneously \cite%
{Marcus_theory,ido5}.

For further illustration, we have calculated the phase-space Wigner distribution
\cite{Schleich} in both examples above and show the snapshots at
intermediated times in Fig. \ref{Flo:fig1}(c,d). In the first example (PLC),
the Wigner distribution at $\tau =100$ exhibits structure characteristic to $n=8$
level of the quantum ladder, as expected from the energy levels occupation
at the same time (see Fig. \ref{Flo:fig1}(a)). In the PAR example at $\tau
=10$ in Fig. \ref{Flo:fig1}(d), the most populated parts of the phase space
are two symmetrically separated resonantly trapped phase space regions of
the parametric oscillator, while the characteristic interference patterns
are seen in the nonresonant regions of phase space. The splitting of the
trapped area in phase-space into two is explained as a pitchfork
bifurcation \cite{Kiselev}.

\emph{Quantum saturation.-} One of the important observables of the
parametrically chirped oscillator is the probability of capture into
resonance. One can define this probability quantum-mechanically as the total
occupation of resonant levels after the sweeping of the modulation frequency
through the linear resonance \cite{ido5}, or classically, as the fraction of
the initial conditions leading to the phase-locked solution \cite{ido3}.
Thus, studying this probability comprises a good framework in discussing the
quantum-classical correspondence in our problem (see a similar discussion in
the direct AR and LC cases in Refs. \cite{ALPHA PRL} and \cite{ido4,ido6},
respectively). In the direct AR scenario for a given temperature, the capture
probability is a smoothed step function of the driving amplitude $%
\varepsilon $ \cite{ido3}. The threshold for capture into resonance in this
case was defined as the value of the driving parameter $\varepsilon _{\text{cr}}$,
yielding $50\%$ capture probability, i.e. $\varepsilon _{\text{cr}}=\varepsilon
(P=0.5)$, while the transition width $\Delta \varepsilon $ was the inverse
slope of $P(\varepsilon )$ at $\varepsilon =\varepsilon _{\text{cr}}$.
$\varepsilon _{\text{cr}}$ is temperature independent while $\Delta
\varepsilon $ scales as $\sqrt{T}$ \cite{ido3,ido5}. Chirped Josephson
circuit experiments revealed that at low temperatures, the AR threshold
width saturates to a finite value, associated with the ground state of the
unperturbed oscillator due to zero point quantum fluctuations \cite%
{ido4,ido5}. This quantum saturation effect was included in the classical AR
theory by introducing an effective temperature $T\rightarrow T_{\text{eff}}=%
\frac{\hbar \omega _{0}}{2k_{\text{B}}}\coth \left( \frac{\hbar \omega _{0}}{2k_{\text{B}}T%
}\right) $, characterizing the thermal state in the Wigner phase-space
representation \cite{ido5}.
\begin{figure}[tp]
\includegraphics[width=6cm]{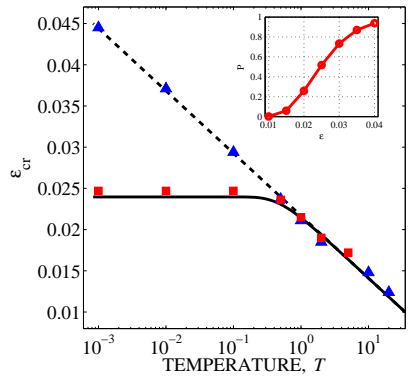}
\caption{(color online) The threshold on the modulation amplitude for resonant
transition in the chirped parametric oscillator versus initial
temperature: the classical simulations (blue $\blacktriangle$),
the theoretical scaling (\protect\ref{eq:epscr}) (dashed line), the Schrodinger simulations (red $\blacksquare$)
exhibiting quantum saturation, and generalized theory  (\ref{eq:epscr_eff}) (solid line). The
insert shows the capture probability $P(\varepsilon)$ at
$T=0.1$ in the quantum simulations (the threshold in this case is $\protect\varepsilon_{\text{cr}}=0.0246$).}
\label{Flo:fig2}
\end{figure}%
In contrast to the direct AR, the low temperature behavior in the PAR has
not been studied previously and is discussed next. Numerical simulations of
Eqs. (\ref{eq:c_n}) show a typical $S$-shape dependence $P(\varepsilon )$, as
in the insert of Fig. \ref{Flo:fig2}, which is similar to that in the direct
AR \cite{ido3}. Therefore, we define the threshold $\varepsilon _{\text{cr}}$ for
the PAR similarly, i.e. $\varepsilon _{\text{cr}}=\varepsilon (P=0.5)$. In the
theory of the classical PAR \cite{lazar90}, $\varepsilon _{\text{cr}}$ was defined
as $\varepsilon (P=0.1)$ due to the asymmetry of $P(\varepsilon )$ for a
fixed initial energy. Nonetheless, for a thermal distribution of initial
conditions,  $P(\varepsilon )$ is symmetric, and $\varepsilon (P=0.5)$ is a
preferable definition. A classical theory of the PAR was developed in \cite%
{lazar90}, showing that the critical driving amplitude scales
logarithmically with the initial action $S_{0}$ of the oscillator, $%
\varepsilon _{\text{cr}}\sim -\ln S_{0}$. Therefore, we expect a similar scaling
with the initial temperature,
\begin{equation}
\varepsilon _{\text{cr}}^{\text{PAR}}(T)=a-b\ln T.  \label{eq:epscr}
\end{equation}%
We verify this scaling in numerical simulations of the classical equation of
motion (\ref{classical}) subject to random, thermally distributed initial
condition \cite{noise_footnote}. The results are shown in Fig. \ref{Flo:fig2}
(blue triangles) for $\alpha =0.0001$, $\beta =0.001$. The logarithmic
scaling is seen in the semi-log figure, while the best-fitted theoretical
parameters in this case are $a=0.0217$ and $b=0.0033$. Note that this
classical scaling predicts an infinitely large driving amplitude for
efficient capture into PAR in the limit of $T\rightarrow 0$. This
singularity is removed if quantum fluctuations are taken into account. To
illustrate this, we solve the Schrodinger equation (\ref{eq:c_n}) associated
with Hamiltonian (\ref{eq:Hamiltonian}) using the parameters of the
aforementioned classical simulations. The anharmonicity here is small, $%
P_{2}=0.075\ll 1$, so many levels are simultaneously coupled at all times
and the dynamics is classical, as illustrated in the insert of Fig. \ref%
{Flo:fig1}(b). The numerical quantum results for $\varepsilon _{\text{cr}}(T)$ are
presented by red squares in Fig. \ref{Flo:fig2}, showing a good agreement
with the classical simulations for temperatures $T>0.5$, but exhibit
saturation of the threshold at low temperatures. This effect is similar to
the quantum saturation of the transition width observed in the direct AR
Josephson junction experiment \cite{ido4}. To include this new
effect in the theory, we replace $T\rightarrow T_{\text{eff}}=\frac{1}{2}%
\coth \left( \frac{1}{T}\right) $ in the classical expression for the
threshold (\ref{eq:epscr}), i.e.
\begin{equation}
\varepsilon _{\text{cr}}^{\text{PAR}}(T)=a-b\ln T_{\text{eff}}.  \label{eq:epscr_eff}
\end{equation}%
This prediction with the coefficients $a,b$ mentioned previously is shown by
a solid line in Fig. \ref{Flo:fig2} and agrees very well with the quantum
numerical results at all temperatures. The replacement $T\rightarrow T_{%
\text{eff}}$ can be explained via the
Wigner phase-space representation. Indeed, the thermal state of the
phase-space representation of a linearized oscillator is
$W(x_{0},p_{0})=(2\pi T_{\text{eff}})^{-1}\exp \left[ -(x_{0}^{2}+p_{0}^{2})/(2T_{\text{eff}})\right]$,
while the quantum Liouville
equation coinsides with the classical Liouville equation in the limit $%
P_{2}\rightarrow 0$. Thus, generally, quantum fluctuations in systems
exhibiting classical dynamics, are taken into account by replacing $%
T\rightarrow T_{\text{eff}}$ in the classical theory. Physically, the
quantum uncertainty principle imposes a limit on the minimal area of the
ground state (at $T=0$) in phase space, while classically, it becomes
infinitesimally small. This imposes a quantum mechanical upper limit on $%
\varepsilon _{\text{cr}}^{\text{PAR}}$, corresponding to the quantum ground state at $%
T\rightarrow 0$ ($T_{\text{eff}}=0.5$). This completes our discussion of the
quantum saturation of the classical PAR at low temperatures. The second
quantum limit, where only few levels are coupled simultaneously at all times
and the dynamics becomes that of the PLC is discussed next.

\begin{figure}[tp]
\includegraphics[width=5.5cm]{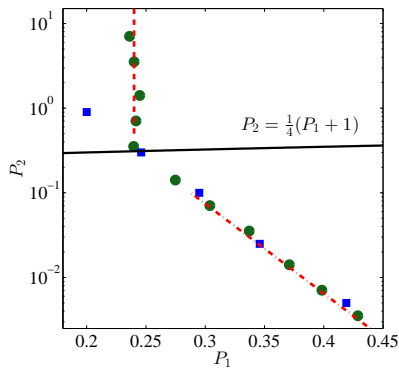}
\caption{(color online) The threshold, $P_{1}^{\text{cr}}$ in $(P_{1},P_{2})$
parameter space: Shrodinger simulations (green $\bullet$), the PLC theory $P_{1,\text{cr}}^{\text{PLC}}=0.237$
(dashed red line), the PAR theory (\protect\ref%
{eq:P1crPAR}) (red dashed dotted line), and the classical simulations
(blue $\blacksquare$). Black solid line separates the classical PAR and the quantum PLC regimes.}
\label{Flo:fig3}
\end{figure}

\emph{Classical-quantum correspondence.-} For studying the transition
between the classical PAR and the quantum PLC regimes, we transform to the
rotating frame as follows. First, we define $C_{n}=c_{n}e^{iE_{n}t}$, and
rewrite Eq. (\ref{eq:c_n}) in the form
\begin{eqnarray}
i\frac{dC_{n}}{dt} &\approx &\frac{\varepsilon }{8}(\sqrt{Q_{n+1}}%
C_{n+2}e^{-i(\omega _{n,n+2}t-\varphi )}  \notag \\
&&+\sqrt{Q_{n-1}}C_{n-2}e^{i(\omega _{n-2,n}t-\varphi )}),  \label{Cn}
\end{eqnarray}%
where $\omega _{n,n+2}=E_{n+2}-E_{n}=2-\gamma (4n+6)$ and we neglect
nonresonant terms (rotating wave approximation). Next, we introduce $%
B_{n}=C_{n}\exp[-i\int \widetilde{\Gamma }_{n}dt]$, where $\widetilde{\Gamma }%
_{n}=\gamma Q_{n} -\frac{1}{2}n\alpha t$ and $\tau =\sqrt{\alpha }t$. Then,
Eq. (\ref{Cn}) can be written as
\begin{equation}
i\frac{dB_{n}}{d\tau } =\Gamma _{n}B_{n}+P_{1}\left(\sqrt{Q_{n+1}}B_{n+2}
+\sqrt{Q_{n-1}}B_{n-2}\right),  \label{Bn}
\end{equation}%
where $\Gamma _{n}=\widetilde{\Gamma }_{n}/\sqrt{\alpha }=\frac{n}{2}%
(P_{2}(n+1)-\tau )$ and $P_{1}=\frac{\varepsilon }{8\sqrt{\alpha }}$, $P_{2}=%
\frac{2\gamma }{\sqrt{\alpha }}$. We have solved the slow Eqs. (\ref{Bn})
numerically, subject to the ground state initial conditions, $B_{n}\left(
\tau _{0}=-10\right) =\delta _{n,0}$, and calculated the resonant capture
probability $P(P_{1})$ for different values of the anharmonicity parameter $%
P_{2}$. The numerics of these slow equations is less time consuming, still
yielding a good agreement with the solutions of the exact equations (\ref{Cn}%
) (not shown). Similar to the definition of $\varepsilon _{\text{cr}}$, we define
the threshold modulation parameter $P_{1}^{\text{cr}}=P_{1}(P=0.5)$ and
show $P_{1}^{\text{cr}}$ in Fig. \ref{Flo:fig3} for different values of the
anharmonicity $P_{2}\in \lbrack 0.0035,7.1]$ (green circles), covering both
the classical PAR and the quantum PLC regimes. We also calculate the
corresponding classical threshold for different $P_{2}$ by solving the
classical equations (\ref{classical}) subject to the initial
\textquotedblleft ground state\textquotedblright {} temperature $T=0.5$. The
resulting $P_{1,\text{class}}^{\text{cr}}(P_{2})$ is shown in Fig. \ref{Flo:fig3}
(blue squares) and agrees well with the quantum calculations in the
classical regime, $P_{2}\ll (P_{1}+1)/4$ (see bellow).

\emph{The theory.-} Next, we develop a theory for $P_{1}^{\text{cr}}(P_{2})$ in
both the quantum PLC and the classical PAR regimes.
In the PLC regime, we assume a sufficiently
large anharmonicity in the quantum PLC regime, such that only two levels are
efficiently coupled as the modulation frequency passes the resonance $\omega
\approx \omega _{n,n+2}$. In this case, the dynamics is that of successive $%
n\rightarrow n+2$ LZ transitions \cite{LZ}, where the avoided resonant
crossing condition occurs when the diagonal terms of the two coupled levels
are equal. Therefore, the time of the $n^{th}$ parametric LZ transition is
derived by equating the diagonal terms in Eq. (\ref{Bn}), i.e., $\Gamma
_{n}=\Gamma _{n+2}$. By solving for $\tau $, one finds that the time of the $%
n\rightarrow n+2$ transition is $\tau _{n}=2nP_{2}$. Therefore, the time
interval between successive parametric LZ transitions in the PLC limit is $%
\Delta \tau =\tau _{n+2}-\tau _{n}=4P_{2}$, in agreement with the example in
Fig. \ref{Flo:fig1}(a) where $P_{2}=10$, so $\Delta \tau =40$. Next, we
write the reduced matrix of the $n\rightarrow n+2$ transition (see Eq. (\ref{Bn}))
\begin{equation}
\left(
\begin{array}{cc}
\frac{1}{2}P_{2}Q_{n}-\frac{1}{2}n\tau & P_{1}\sqrt{Q_{n+1}} \\
P_{1}\sqrt{Q_{n+1}} & \frac{1}{2}P_{2}Q_{n+2}-\frac{1}{2}(n+2)\tau%
\end{array}\right).  \label{TrMatrix}
\end{equation}%
The typical time of the $0\rightarrow 2$ LZ transition in the adiabatic
limit is $\Delta \tau _{\text{LZ}}=P_{1}$ and unity in the nonadiabatic
limit \cite{ido3,Finite LZ}. Therefore, the theoretical condition for well
separated successive LZ transitions (i.e., PLC), $\Delta \tau \gg \Delta
\tau _{\text{LZ}}$, yields $P_{2}\gg \frac{1}{4}(P_{1}+1)$. In this limit,
the probability of each $n\rightarrow n+2$ transition is given by the LZ
formula \cite{LZ},
\begin{equation}
P_{n\rightarrow n+2}=1-e^{-2\pi P_{1}^{2}Q_{n+1}}
\end{equation}%
and the total probability for capture into the PLC starting from
the ground state is $P_{total}=\prod_{n=0}^{\infty }P_{n\rightarrow n+2}$.
Solving $P_{total}=0.5$ for $P_{1}$, yields the threshold for capture into
the PLC, $P_{1,\text{cr}}^{\text{PLC}}=0.237$, where only two first terms in the product
are needed for less than $1\%$ accuracy. This theoretical prediction is
shown in Fig. \ref{Flo:fig3} by a dashed red line and agrees well with the
numerical simulations in the quantum PLC regime. In addition, the
theoretical separator $P_{2}=\frac{1}{4}(P_{1}+1)$ (black solid line)
predicts correctly the location of the transition between the quantum PLC
and the classical PAR regimes.

Finally, in the PAR  regime
we write $x=a\cos\theta$, define the slow phase-mismatch $\phi =2\theta -\varphi$ and
the rescaled amplitude $A=\sqrt{P_{2}}a$, employ the single resonance
approximation, and average Eq. (\ref{classical})~over the fast phase of the
oscillator. This yields \cite{Khain}
\begin{equation}
\frac{dA}{d\tau }=2P_{1}A\sin \phi ;\quad \frac{d\phi }{d\tau }=A^{2}-2\tau
+4P_{1}\cos \phi ,
\end{equation}%
where, as before, $P_{1}=\frac{\varepsilon }{8\sqrt{\alpha }}$, $P_{2}=\frac{%
3\beta }{4\sqrt{\alpha }}$, $\tau =\sqrt{\alpha }t$, and the initial thermal
distribution is $f(A_{0})=\sigma ^{-2}A_{0}\exp\left[-A_{0}^{2}/2\sigma ^{2}\right]$,
$\sigma ^{2}=0.5P_{2}T$.
Then, the generalized expression for the threshold becomes
\begin{equation}
P_{1,\text{cr}}^{\text{PAR}}(T)=\kappa _{0}-\kappa _{1}\ln (P_{2}T_{\text{eff}})
\label{eq:P1crPAR}
\end{equation}%
where $\kappa _{0}=0.165$ and $\kappa _{1}=0.41$, are obtained by comparing
Eqs. (\ref{eq:epscr_eff}) and (\ref{eq:P1crPAR}). The numerical results in
Fig. \ref{Flo:fig3} agree with this prediction for $T_{\text{eff}}=0.5$ (red
dashed-dotted line) in the classical $P_{2}\gg \frac{1}{4}(P_{1}+1)$ regime.

In summary, we have studied the problem of passage through the parametric
resonance in a quantum anharmonic oscillator and identified the quantum
counterpart of the classical PAR, i.e., the quantum PLC. We have developed a
theory of the PLC and the PAR for thermal initial conditions
and found the threshold of capture into resonance
in both regimes. We have also studied the transition from the PLC to the
classical PAR and illustrated both the classical and the quantum resonant
dynamics in phase-space by the Wigner distribution. In addition, we
have identified the effect of quantum saturation of the threshold for
parametrically resonant transition at small temperatures due to zero-point
fluctuations. The saturation defines the maximum modulation amplitude needed
for efficient PAR excitation. These results pave the way for using the PLC
and the PAR as robust control tools in quantum electronic or optical systems
in such applications as quantum communication and computing. It also seems
interesting to extend the study of the PLC to quantum systems of many
degrees of freedom, such as complex molecules and coupled qbits for
controlling different interacting degrees of freedom. Supported by the
Israel Science Foundation (Grant No. 451/10).


\begin{thebibliography}{99}
\bibitem{noise_footnote} Similar to the direct AR, the coupling of the
driven oscillator to a thermal environment can be modeled by a deterministic
oscillator (with no external noise) but with thermally distributed initial
conditions, provided sufficiently slow diffusive time as comparison to the
sweeping time of driving frequency through the linear resonance \cite{ido3}.
\end{thebibliography}
\end{document}